\documentclass[12pt]{article}

\usepackage{wrapfig}           
\usepackage{graphicx}          
\usepackage{dcolumn}           
\usepackage{bm}                

\title{Einstein's static universe }

\author{Domingos Soares \\ Departamento de F\'{\i}sica, 
Universidade Federal de Minas Gerais \\C.P. 702,   
30123-970,  Belo Horizonte, Brazil} 

\date{\today}

\begin{document}

\maketitle


\begin{abstract}
Einstein's static model is the first relativistic cosmological 
model. The model is static, finite and of spherical spatial 
symmetry. I use the solution of Einstein's field equations in a 
homogeneous and isotropic universe --- Friedmann's equation ---  
to calculate the radius of curvature of the model (also known as 
\emph{Einstein's universe}). Furthermore, I show, using a 
Newtonian analogy, the model's mostly known feature, namely, its 
instability under small perturbations on the state of equilibrium. 
\end{abstract}


\section{Introduction}

In 1917, therefore, less than a hundred years ago, Albert Einstein 
(1879-1955) put forward the first relativistic cosmological model, i.e., 
a model based in the General Relativity Theory (GRT), that he had just 
finished (\cite{desou}, chap. 8, \cite{caos}, chap. 27, \cite{harr}, 
chap. 14, \cite{waga}, section 2). 

The model, nowadays considered as surpassed, represented a most 
profitable seed of a series of theoretical studies which had the aim 
of understanding the general structure of the universe, both in 
space and time. Einstein's model is the starting point of 
relativistic cosmology. The model is static, with positive 
spatial curvature (closed), therefore, spatially bound --- 
in other words, finite. It was static because this was the general 
view of the real universe at the time, and finite, because 
being so it avoided the necessity of infinite quantities 
as boundary conditions, an undesirable feature in any 
physical theory. It is worthwhile mentioning that, in 1917, 
the hypothesis of a static universe 
was quite reasonable. The observations by Edwin Hubble (1889-1953),  
that would be consistent with non static solutions, had not yet 
been realized (see detailed discussion on this issue in \cite{soar}). 

In order to achieve those characteristics it was necessary to 
counterbalance the attractive effects of gravity. Einstein 
introduced a constant in his field equations --- the now famous 
\emph{cosmological constant} ---, as a repulsive term, at the right 
amount, to make possible the sort of solution he needed. Besides being in 
accordance with the common views of his epoch, of a static universe, 
Einstein aimed also to justify the ideas of the Austrian physicist and 
philosopher E. Mach (1838-1916) regarding the genesis of the property of 
inertia. According to Mach, the inertial mass of any body is due to the 
influence of the universe as a whole. Einstein agreed with such an idea and 
believed that his model connected local properties --- the mass --- with global properties --- the cosmological constant (\cite{harr}, p. 272). Incidentally, 
later on, Einstein's enthusiasm with respect to Mach's principle diminished 
and finally disappeared completely (see, for example, \cite{pais}, p. 287). 

Almost immediately after Einstein's proposition, the Dutch W. de 
Sitter (1872-1934), the Russian A. Friedmann (1888-1925) and 
the Belgian G. Lema\^itre (1894-1966) came up 
with alternative models to Einstein's static model, also based in GRT. The 
models of W. de Sitter, A. Friedmann and G. Lema\^itre have a peculiarity 
that does not exist in Einstein's model: they represent expanding universes. 
The light emitted by any galaxy arrives at an observer in a distant galaxy 
with its wavelength shifted towards  the red, i.e., redshifted. In other words, 
light arrives with a wavelength larger than the wavelength at emission. Such 
a property does not exist in Einstein's model because it represents a static 
universe. W. de Sitter's model, on the other hand, has a feature that contributed 
to lessen its importance: it represents a universe completely without matter and 
radiation, where galaxies are interpreted as test particles immersed in the 
expanding space-time. It shares with Einstein's model the inclusion of a 
cosmological constant. As mentioned above, and as it will be explained in the next 
section, in Einstein's model the cosmological constant is responsible for 
the tendency to expansion that is exactly matched by the tendency to attraction due 
to matter and radiation. The latter do not exist in W. de Sitter's model, therefore, 
this model shows only expansion. 

It was soon realized then that Einstein's model was unstable for small 
perturbations to the state of equilibrium. And finally, the British 
astrophysicist Arthur Eddington (1882-1944) showed definitely that the model 
was unstable \cite{edd}, lending capital doubts on its viability.

In the next section, I use Friedmann's equation, modified by the 
introduction of the cosmological constant, to calculate the radius of 
Einstein's universe. Next, I analyze the potential energy of a 
Newtonian analogy to show that this universe is in a state of 
unstable equilibrium. I finish with a discussion of Einstein's 
self-criticism about his first model of the universe.

\section{The radius of Einstein's static universe}

Friedmann's equation is a general solution of the field equations of GRT 
under the constraints of a homogeneous and isotropic fluid (see 
\cite{viso}). Einstein's field equations can be expressed in a synthetic 
form by means of the tensor formalism. Thus one has on the left side of the 
equation the energy-momentum tensor and on the right side the curvature tensor, 
which represents the system's space-time characteristics (see, for example, 
eq. 3.6 in \cite{desou}). In a simple manner, one can say that that the 
mass and energy contents of the system \emph{say} to space-time how to curve. 
Curved space-time \emph{says} then to a test particle in it how to move. 

The TRG field equations are, in fact, a system of non linear differential equations 
of extremely difficult solution. However, for a fluid that is homogeneous 
--- same density everywhere --- and isotropic --- same properties at all 
directions ---, as mentioned above, the system of equations is simplified 
allowing for analytical solutions, such as, for example, Friedmann's equation. 

Friedmann's equation has on the left-hand side the energy terms and on the 
right side the curvature term. It is written, in terms of the curvature 
constant of the system, $K_\circ$, as (\cite{desou}, eq. 2.19):
\begin{equation}
\label{eq:fried}
\left(\frac{dR}{dt}\right)^2 - \frac{8\pi G}{3}\rho R^2 = 
-K_\circ c^2, 
\end{equation}

\noindent where $R$ is the scale factor and $\rho$ is the total density  
in $R(t)$. $G$ is the universal gravitational constant and $c$ is the speed of 
light in vacuum. The density $\rho$ varies with time and its present observed 
value is approximately $10^{-30}$ g/cm$^3$.   
The curvature constant is, for a closed spherical universe, 
$K_\circ=+1/{\cal{R}}^2$, and $\cal{R}$ is the radius of curvature of the 
spherical space. For a critical (or flat) model $\cal{R}\rightarrow \infty$, 
and, therefore, $K_\circ=0$. The open universe has an imaginary radius of 
curvature, meaning that it has a negative constant of curvature 
$K_\circ=-1/{\cal{R}}^2$ (hyperbolic space).   

This equation was obtained, for the first time, by the Russian 
Alexander Friedmann in 1922. It is used here in the discussion 
of Einstein's model because it makes much more simpler the derivation of both 
the radius of the universe and the investigation of the model's stability. 
Historically, though, that was not the way followed by Einstein, because 
his model was devised in 1917.
 
Eq.~\ref{eq:fried} can be modified, without violating GRT, by adding a 
constant, conveniently expressed as $1/3\Lambda c^2$, on the left-hand side 
of the equation. This additional term can also be considered as a  
density term $\rho_\Lambda= \Lambda c^2/8\pi G$. Hence, one has

\begin{equation}
\label{eq:fried0}
\left(\frac{dR}{dt}\right)^2 - \left(\frac{8\pi G}{3}\rho   
+\frac{1}{3}\Lambda c^2\right)R^2 = -K_\circ c^2, ~~{\rm or}
\end{equation}
\begin{equation}
\label{eq:fried00}
\left(\frac{dR}{dt}\right)^2 - \frac{8\pi G}{3}\left(\rho   
+\rho_\Lambda\right)R^2 = -K_\circ c^2, ~~{\rm and~finally} 
\end{equation}
\begin{equation}
\label{eq:fried1}
\left(\frac{dR}{dt}\right)^2 -\frac{8\pi G}{3}\frac{\rho_\circ}{R} 
-\frac{1}{3}\Lambda c^2R^2 = -K_\circ c^2, 
\end{equation}

\noindent with $\rho(t)R(t)^3=\rho(t_\circ)R(t_\circ)^3$, 
or, $\rho R^3=\rho_\circ$, where $t_\circ$ is the presente time, 
$\rho_\circ$ is the density in $t_\circ$ and $R(t_\circ)$ is, 
conventionally, set to 1. The transformation $\rho R^3=\rho_\circ$ 
is nothing more than the expression of mass conservation in an 
evolving universe (density times volume, i.e., mass, is constant), 
which is valid also, of course, for the special case of a static 
universe. 

The cosmological constant $\Lambda$ has the physical dimension of 
1/length$^2$. According to the cosmologist Wolfgang Rindler 
(\cite{rind}, p. 303), \emph{``The $\Lambda$ term [\dots] seems to 
be here to stay; it belongs to the field equations much as an additive 
constant belongs to an indefinite integral''}. While, mathematically, 
the cosmological constant preserves the validity of GRT's field 
equations, physically, it leads to multiple possible consequences 
in the behavior of model universes.

Differentiating eq. \ref{eq:fried1} with respect to time, yields:

\begin{equation}
\label{eq:fried2}
2\dot{R}\ddot{R} +\frac{8\pi G}{3}\frac{\rho_\circ}{R^2}\dot{R}  
-\frac{2}{3}\Lambda c^2R\dot{R} = 0, 
\end{equation}

\noindent which can be simplified to 

\begin{equation}
\label{eq:fried3}
\ddot{R} +\frac{4\pi G}{3}\frac{\rho_\circ}{R^2}  
-\frac{1}{3}\Lambda c^2R = 0. 
\end{equation}

With constant $R$, eq. \ref{eq:fried3} clearly shows that $\Lambda$ 
can be fine-tuned to yield $\ddot{R}=0$, thus, implying a static solution, 
which was precisely Einstein's desire.  

As mentioned above, in Einstein's static model the constant of 
curvature is $K_\circ=1/{\cal{R}}^2$. With this and making the scale 
factor $R\equiv R_E=1$ in eqs. \ref{eq:fried1} and \ref{eq:fried3}, 
one gets the following two relations:

\begin{equation}
\label{eq:fried4}
-\frac{8\pi G}{3}\rho_\circ - \frac{1}{3}\Lambda c^2 = 
-\frac{c^2}{{\cal{R}}^2} 
\end{equation}
\begin{equation}
\label{eq:fried5}
\frac{4\pi G}{3}\rho_\circ - \frac{1}{3}\Lambda c^2 = 0. 
\end{equation}

Inserting eq. \ref{eq:fried5} in eq. \ref{eq:fried4} gives 

\begin{equation}
\label{eq:fried6}
4\pi G\rho_\circ = \frac{c^2}{{\cal{R}}^2} 
\end{equation}

\noindent or  

\begin{equation}
\label{eq:fried7}
{\cal{R}} = \frac{c}{\sqrt{4\pi G\rho_\circ}}, 
\end{equation}

\noindent that is the radius of curvature of Einstein's static 
universe.

What is its numerical value? For the sake of illustration, let 
us take $\rho_\circ=3H_\circ^2/8\pi G$, namely, the density of 
Friedmann's critical model, also known as the Einstein-de Sitter model. 
Here, $H_\circ$ is Hubble's constant (see \cite{desou} and \cite{caos} 
for more details about such a model). Then, one gets ${\cal{R}} = 
\sqrt{2/3}(c/H_\circ) = 3.4$ Gpc $= 11$ Gly, with $H_\circ=$ 
72 km s$^{-1}$Mpc (cf. \cite{freed}).

It is worthwhile stressing that the above calculation of ${\cal{R}}$ 
is just illustrative, having no real physical meaning. In the days 
when Einstein put forward his static model, the value used for the 
density was the observed density, which, coincidently, in order 
of magnitude, did not differ from the value exemplified above.

\section{Study of stability}

As shown in the beginning of the preceding section, Friedmann's 
equation (eq. \ref{eq:fried}) has the energy terms on its lef-hand side 
and the curvature term --- which is constant --- on its right side. 
A Newtonian analogy may be built from eq. \ref{eq:fried1}. Such an 
equation represents the conservation of total energy, applied to the 
cosmic fluid. We shall use, for the analogy, Friedmann's equation modified 
with the addition of the cosmological constant, in the form of 
eq. \ref{eq:fried1}.

The right-hand side term represents the total energy of the 
system --- negative, i.e., a bound system, as is the case in 
Einstein's model. The first term on the left-hand 
side represents the kinetic energy of the cosmic fluid element, the 
second term its gravitational potential energy and the third term 
--- of the $-1/2kx^2$ kind --- represents a sort of repulsive 
``elastic'' cosmic potential energy. This last term, in Friedmann's 
equation, could be thought of as an intrinsic stress in the 
space-time tissue, quantified by the cosmological constant. In the 
analogous Newtonian construction, it is regarded as an elastic potential 
energy of a string, with the important difference of being a \emph{negative} 
energy term. The second term will be, then, represented by $U_G=-1/R$ and the 
third one by $U_\Lambda=-1/2R^2$. 

The radial forces related to these potential energies can be calculated by  
$F=-dU/dR$, yielding $F_G=-1/R^2$ and $F_\Lambda=+R$, the first, an attractive 
force --- driven by gravitation --- and the second one, a repulsive force --- 
driven by the cosmic ``elasticity'', much like the same as a rubber sheet 
would do --- the space-time tissue --- upon a body that rests on it. 
These two forces balance exactly in Einstein's static universe.

\begin{figure}[ht]
\centering
\includegraphics[width=10cm,height=10cm]{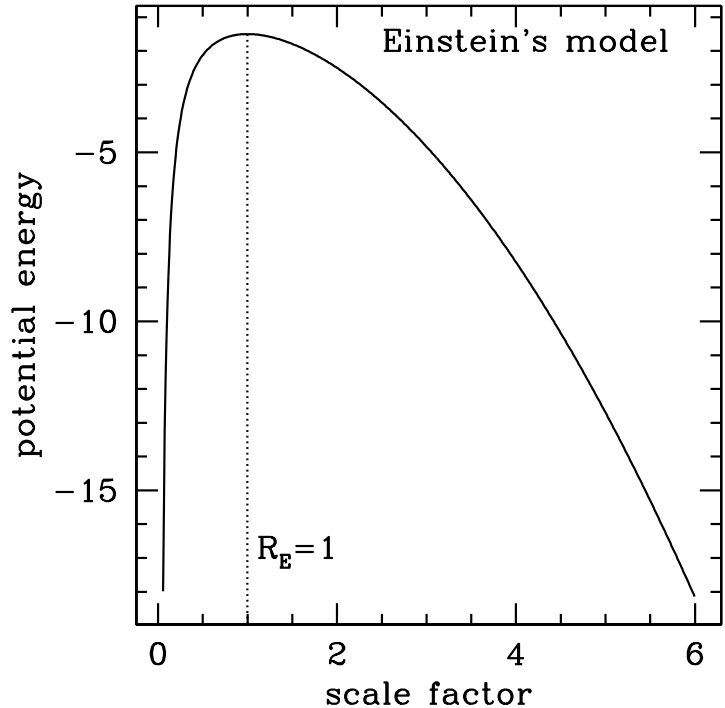}
\caption{\label{fig:unvstt} {\footnotesize $\Lambda$-shaped diagram: 
the potential energy --- in 
arbitrary units --- for the Newtonian analogy of Einstein's static 
model. Notice that the equilibrium at $R=R_E$ is an unstable one. 
Any small perturbation at $R_E$ makes either the universe 
to collapse or diverge to $R\rightarrow\infty$.}
}
\end{figure}

Therefore, the conservation of energy in the Newtonian analogy may be 
written as 

\begin{equation}
\label{eq:newt0}
\frac{1}{2}mv^2 +U_G + U_\Lambda = -E 
\end{equation}
\begin{equation}
\label{eq:newt1}
\frac{1}{2}mv^2 -\frac{1}{R}-\frac{1}{2}R^2 = -E, 
\end{equation}

\noindent where $-E<0$ is the system's total energy. Fig. \ref{fig:unvstt} 
shows the total potential energy function $U=U_G+U_\Lambda$. It is quite 
apparent that the point of equilibrium  --- $F=-dU/dR=0$ --- represents an 
unstable equilibrium. Precisely what we would like to show. 

\section{Final remarks}

Soon after Einstein put forward his cosmological model, two almost 
simultaneous events, in the beginning of the 1920s, changed in a dramatic 
way the scientific view of the universe. One of them was the discovery 
of the systematics exhibited by the spectral shifts of the radiation 
emitted by extragalactic nebulae, undertaken by Edwin Hubble. The other 
one was the discovery of new solutions of Einstein's field equations, 
by Friedmann (see eq. \ref{eq:fried}, above, used in section 2), that 
implied in dynamical models. The universe could be either in expansion 
or in contraction, and the first possibility was consistent with Hubble's 
observations. There was not anymore the necessity of a static model.

It is rather well known Einstein's reaction to these great events.

The renowned theoretical physicist John Archibald Wheeler (1911-2008) 
tell us that, once, as a young scientist, he went along with Einstein 
and George Gamow (1904-1968), in the Institute of Advanced Studies, in 
Princeton, when he heard Einstein confess to Gamow that the 
cosmological constant had been \emph{``the biggest blunder of my life''} 
(cf. \cite{tawh}, p. G-11). 

Obviously, Einstein was not a fool, and the inclusion of $\Lambda$ in his 
field equations, definitely, was not a blunder at all. It increased in 
a substantial way the applicability of GRT, without causing damages from 
the formal point of view, as mentioned in section 2.

In fact --- and it is something that probably Einstein did not want 
to recognize ---, his real big blunder was to put forward a model 
that was clearly unstable. The fact that he was not aware of that is 
that causes a big surprise. As we saw, in section 3, a simple analogous 
in classical reasoning makes clear such a very serious failure.


\end{document}